# Algebraic Nilsson cranking model and its prediction for $^{20}$Ne

P. Gulshani[1] and A. Lahbas[2]

[1]3313 Fenwick Crescent, Mississauga, Ontario, Canada L5L 5N1, parviz.gulshani@outlook.com

[2]ESMaR, Faculty of Science, Mohammed V University in Rabat, Morocco, a.lahbas@um5r.ac.ma

Corresponding author: P. Gulshani

**Abstract**

Previously, Nilsson-Ragnarsson solved numerically the conventional cranking model (CCRM3) with deformed oscillator potential and spin-orbit interaction (refer to as "Nilsson CCRM3") and predicted the $^{20}$Ne ground-state rotational-band yrast energies at the angular momenta $I$=2,4,6, and 8. However, CCRM3 is semi-classical and phenomenological and breaks a number of nuclear symmetries. Recently, we developed, starting from the nuclear Schrodinger equation, a microscopic, quantal, self-consistent cranking model (MSCRM3), where, among other features, the angular velocity is microscopic derived. We solved algebraically the MSCRM3 equations for the pure oscillator potential and used the model to predict energies and rotation types in $^{20}$Ne. Some interesting results were obtained, such as the quenching or the transition of planar rotation to a uniaxial rotation thereby reducing the excitation energy at $I$=8 in $^{20}$Ne. In this article, we use the algebraic method developed in our MSCRM3 analysis to solve iteratively the self-consistent Nilsson-CCRM3 Schrodinger equation. The application of this algebraic Nilsson-CCRM3 model to the $^{20}$Ne nucleus predicts ground-state rotational-band excitation energies at the angular momenta $I$=2,4,6, and 8 that are in a much better agreement with the measured energies than those predicted earlier by Nilsson-Ragnarsson using a numerical solution method. This agreement and the predicted excited-state energies at $I$ =4 and 8 that vary periodically with the iteration steps (because of single-particle level crossings) provide a possible explanation for the measured lower yrast-state energies at $I$ =4 and 8 relative to those at the $I$ =2 and 6, somewhat resembling the quenching of planar rotation at $I$ =8 mentioned above. This better agreement may also provide a further indication of the weakness of the pairing correlations in $^{20}$Ne.

## 1. Introduction

The conventional cranking model for triaxial collective rotation (CCRM3) is described by the Schrodinger equation:

$$\hat{H}_{cr}|\Phi_{cr}\rangle \equiv \left(\hat{H}_o - \vec{\Omega}\cdot\vec{\hat{J}}\right)|\Phi_{cr}\rangle \equiv E_{cr}|\Phi_{cr}\rangle \qquad (1)$$

where $\hat{H}_o$ is the rotational and time-reversal invariant nuclear Hamiltonian, $\vec{\hat{J}}$ is the total angular-momentum-vector operator, and $\vec{\Omega}$ is a constant angular-velocity vector. The CCRM3 model, with a rotationally non-invariant $\hat{H}_o$ (for example, the Nilsson's oscillator model) has frequently been used to predict uniaxial and triaxial rotation properties in deformed nuclei [1-11]. However, it is recognized [2,3,12] that the model is semi-classical and phenomenological because it uses a constant adjustable angular velocity $\vec{\Omega}$ parameter. Therefore, Eq. (1) is time-reversal and $D_2$, and signature non-invariant, ignores the interaction and feedbacks between the angualr velocity and angulr momentum, and the relation of Eq. (1) to the time-reversal and rotation invariant nuclear Schrodinger equation is not clear. There have been many revealing investigations [3,6,12-18] to derive the model from first principles (i.e., from rotation-invariant nuclear Schrodinger equation) using various methods, and discover the implicit model assumptions and approximations. The major difficulty in deriving a tractable microscopic quantal cranking model has been the strong coupling between the rotation and intrinsic motions.

In previous articles[1], we derived a microscopic, quantal, self-consistent, time-reversal and $D_2$ invariant cranking model Schrodinger equation for triaxial rotation (MSCRM3), where the angular velocity is microscopically determined, and containing no free parameters. Except for the microscopic angular velocity and normally small square of the angular-momentum and two-body-interaction residual terms and other correction terms, the MSCRM3 and CCRM3 Schrodinger equations are identical in form. These residual and correction terms in MSCRM3 may become significant under certain conditions such as in wobbly and tilted-axis rotations and for triaxial rotations at high spins and in back-bending regions where the angular velocity and angular momentum undergo large changes. Indeed, the square of the angular momentum residual term is responsible for restoring the rotational symmetry that is broken in the HF mean-field approximation. When the residual of the square of the angular momentum is included in the HF mean-field Hamiltonian in MSCRM3, it becomes feasible to link the deformed HF mean-field nucleon wavefunction to the isotropic nuclear angular-momentum eigenstate (without using an angular-momentum projection methodology), particularly in the back-bending region. The $D_2$ and signature invariances render the MSCRM3 wavefunction a linear superposition of either even or odd angular-momentum eigenstates, in contrast to a CCRM3 wavefunction that is a superposition of even and odd angular-momentum eigenstates. This result resolves a conceptual problem in comparing predicted and measured nuclear spectra identified in [19]. Time-reversal invariance in MSCRM3 renders its states and its time-reversed companion states degenerate possibly impacting consequences of nucleon pairing interaction.

In previous articles (refer to footnote 1) we determined, for the simple triaxial harmonic oscillator potential, an algebraic solution of MSCRM3 self-consistent equations, and some rotation properties in light nuclei $^{20}$Ne, $^{24}$Mg, and $^{28}$Si. Comparison of these properties with those predicted by CCRM3 revealed some interesting differences. Some reviewers have suggested that the model potential energy should include spin-orbit interaction and pairing correlations.

In this article, we solve algebraically the CCRM3 self-consistent equations for triaxial oscillator that includes spin-orbit interaction (refer to as "Nilsson single-particle potential"), and apply the model to $^{20}$Ne, noting that, in $^{20}$Ne, pairing correlations are insignificant [4,7] and hence they are ignored in the analysis in this article. We compare the predicted nuclear ground-state deformations and yrast-state excitation energies with those measured and presented in [4,9], and provide comments on the differences among the results.

In Section 2, we present the Nilsson-CCRM3 equations for triaxial harmonic oscillator potential that includes spin-orbit interaction [4,9] but neglects pairing correlations since they are insignificant in $^{20}$Ne [4,7]. We solve the Nilsson-CCRM3 Schrodinger equation, subject to self-consistency and constant-volume conditions, using the algebraic method described in [refer to footnote 1] and the similarity between angular-moment cranking term and the spin-orbit interaction term. Section 3 presents the deformations and ground-state rotational-band yrast energies predicted by the model for $^{20}$Ne and compare them with those predicted in [4] and [refer to footnote 1]. Differences between the two sets of model predictions in this article and in [4] are discussed. Section 4 presents a summary of the predicted results, conclusions, and related future analysis plan.

## 2. Algebraic solution of Nilsson-CCRM3 equations

The Nilsson single-particle Hamiltonian is given in [4] by:

$$h_N = \frac{p^2}{2M} + \frac{M}{2}\left(\omega_1^2 x^2 + \omega_2^2 y^2 + \omega_3^2 z^2\right) - \frac{c}{\hbar^2}\vec{l}\cdot\vec{s} - \frac{D}{\hbar^2}\left(l^2 - \langle l^2 \rangle_N\right) \qquad (2)$$

In Eq (2), $\vec{l}$ and $\vec{s}$ are respectively the single-particle angular-momentum and spin vector operators. The CCRM3 single-particle Hamiltonian is then given by (where pairing correlations are ignored, as in [4,9], because they are not significant in s-d shell nuclei):

$$h_{Ncr} = \frac{p^2}{2M} + \frac{M}{2}\left(\omega_1^2 x^2 + \omega_2^2 y^2 + \omega_3^2 z^2\right) - \frac{c}{\hbar^2}\vec{l}\cdot\vec{s} - \frac{D}{\hbar^2}\left(l^2 - \langle l^2 \rangle_N\right) - \vec{\Omega}_c \cdot \vec{j}, \qquad (3)$$

---
[1] P. Gulshani, arXiv:2204.14207 (nucl-th), 29 April, 2022; P. Gulshani and A. Lahbas, arXiv:2308.10026 (nucl-th), 19 August, 2023

where:

$$\vec{j} \equiv \vec{l} + \vec{s}, \quad c \equiv 2\hbar \overset{o}{\omega}_o \kappa, \quad D \equiv \hbar \overset{o}{\omega}_o \kappa\mu, \quad \langle l^2 \rangle_N = \frac{1}{2} N(N+3) \quad (4)$$

$\vec{\Omega}_c$ is a constant cranking angular velocity, and $\overset{o}{\omega}_o$ is the isotropic-oscillator frequency given in [4] by

$\hbar \overset{o}{\omega}_o = P \cdot A^{-1/3}$, $P = 35.43$ for light nuclei, and 41 for heavy nuclei. In Eq (4), $\kappa = 0.08$ and $D = 0$ for light (s-d shell) nuclei, and hence we ignore the 4th term on the right-hand sides of Eqs (2) and (3). In the solution method we are using, it is convenient to re-express Eq. (3) as follows:

$$h_{Ncr} = \frac{p^2}{2M} + \frac{M}{2}\left(\omega_1^2 x^2 + \omega_2^2 y^2 + \omega_3^2 z^2\right) - \vec{l} \cdot \vec{\Omega} - \vec{\Omega}_c \cdot \vec{s}, \quad (5)$$

where:

$$\vec{\Omega} \equiv \vec{\Omega}_c + \frac{c}{\hbar^2}\vec{s} = \vec{\Omega}_c + 2\kappa\vec{s}, \quad (6)$$

where we have used Eq.(4) for $c$ and have divided all quantities in Eq (6) by $\hbar \overset{o}{\omega}_o$, and $\vec{s}$ by $\hbar$. $\vec{s}$ commutes with all the variables in Eq. (5) except with itself. Hence, its commutation relation with $h_{Ncr}$ results in a term related to the precession of $\vec{s}$ about the angular-momentum vector operator $\vec{l}$, and this term is very small. Therefore, for now we ignore the last term in Eq. (5) and replace $\vec{s}$ in Eq. (6) by its expectation value $\langle \vec{s} \rangle$, and Eq. (6) becomes:

$$\vec{\Omega} = \vec{\Omega}_c + 2\kappa\langle \vec{s} \rangle \equiv \vec{\Omega}_c + \vec{\Omega}_s \quad (7)$$

We solve the cranked single-particle Schrodinger equation for $h_{Ncr}$ in Eq (5), without the 4th term, using the Heisenberg equations of motion for the particle-coordinate radius vector $\vec{x}$ prescribed in Eqs (8) and (9):

$$\vec{x}(t) = e^{i\hat{h}_{Ncr} t/\hbar} \vec{x} e^{-i\hat{h}_{Ncr} t/\hbar} \quad (8)$$

$$M\frac{d^2}{dt^2}x_k(t) = -M\omega_k^2 x_k(t) - 2M\left(\vec{\Omega} \times \frac{d}{dt}\vec{x}(t)\right)_k - M\left(\vec{\Omega} \times \left(\vec{\Omega} \times \vec{x}(t)\right)\right)_k \quad (9)$$

where $k=1,2,3$. We seek a periodic motion for $\vec{x}(t)$ with a frequency $\alpha$ as follows:

$$\vec{x}(t) = \vec{x}_o \cdot e^{i\alpha t} \quad (10)$$

We determine the three normal mode frequencies $\alpha_k$ ($k=1,2,3$) associated with the Hamiltonian $h_{Ncr}$ in Eq (5) by inserting Eq (10) into Eq (9) to obtain the following cubic characteristic equation for the three normal-mode frequencies $\alpha$:

$$\alpha^6 - A_o \alpha^4 + B_o \alpha^2 - C_o = 0 \quad (11)$$

where $A_o$, $B_o$, and $C_o$ are polynomial functions of $\omega_k^2$ and $\Omega_k^2$. The resulting three decoupled harmonic-oscillator single-particle eigenstates $|v\rangle = |N, n_1, n_2, n_3\rangle = |n_1\rangle|n_2\rangle|n_3\rangle$ and energy eigenvalues $e_v$ are then given in terms the three oscillator phonon quantum numbers $n_1$, $n_2$, $n_3$ along the three coordinate axes $x$, $y$, and $z$, and the total quantum number $N = n_1 + n_2 + n_3$:

$$e_v = \hbar \sum_{k=1}^{3} \alpha_{v,k} \Sigma_{v,k}, \quad \Sigma_{v,k} \equiv n_{v,k} + 1/2 \quad (12)$$

The cranked nucleus intrinsic (i.e., in the rotating frame of reference) energy is then given by:

$$E_{int} = \sum_{v=1}^{v_f} \langle v | h_{Ncr} | v \rangle = \sum_{v=1}^{v_f} e_v \quad (13)$$

where the subscript $f$ refers to the highest particle-occupied (Fermi) energy level. The rotational-band excited-state and excitation energies, respectively $E_I$ and $\Delta E_I$, at the desired angular momentum quantum number $I$ are then given by:

$$E_I \equiv \sum_{\nu=1}^{\nu_f}\langle\nu|h_N|\nu\rangle \equiv \sum_{\nu=1}^{\nu_f}\langle\nu|(h_{Ncr}+\vec{\Omega}_c\cdot\vec{j})|\nu\rangle = E_{int} + \vec{\Omega}_c\cdot\sum_{\nu=1}^{\nu_f}\langle\nu|\vec{l}|\nu\rangle + \vec{\Omega}_c\cdot\langle\vec{s}\rangle \tag{14}$$

$$\Delta E_I \equiv E_I - E_{I=0} \tag{15}$$

We determine $\vec{\Omega}_c$ and $\vec{\Omega}_s$ in Eq (7) by expressing them in polar coordinates as follows:

$$\vec{\Omega}_c = \Omega_{co}\cdot(\sin\theta\cdot\cos\phi, \sin\theta\cdot\sin\phi, \cos\theta) \tag{16}$$

$$\vec{\Omega}_s = \Omega_{so}\cdot(\sin\theta\cdot\cos\phi, \sin\theta\cdot\sin\phi, \cos\theta), \tag{17}$$

where $\Omega_{co}$ in Eq (16) is the magnitude of the cranking angular velocity, which is varied until the predicted angular momentum $\sum_{\nu=1}^{\nu_f}\sqrt{\langle\nu|(l_1^2+l_2^2+l_3^2)|\nu\rangle}$ equals the desired angular momentum $\hbar I$. In Eq (17), the magnitude of the spin is set to its normal value, namely:

$$\Omega_{so} = 2\kappa s_o = 2\kappa\sqrt{\frac{1}{2}\cdot(\frac{1}{2}+1)} = \kappa\sqrt{3} \tag{18}$$

In Eqs (16) and (17), the polar angles $\theta$ and $\phi$ are determined using Feynman's theorem [20] and requiring them to minimize the intrinsic energy $E_{int}$ in Eqs (5) and (12), as was done in [refer to footnote 1]. We then obtain:

$$\tan\phi = \langle l_2\rangle/l_1, \qquad \tan\theta = \sqrt{1+\tan^2\phi}\cdot\langle l_1\rangle/l_3 \tag{19}$$

The expectation of $\vec{l}$ is similarly obtained by minimizing $E_{int}$ with respect to $\vec{\Omega}$. Constant nuclear volume condition and the condition of self-consistency between the shapes of nuclear density and mean-field (i.e., harmonic oscillator) potential energy (which implies minimization of $E_{int}$ with respect to the oscillator-potential frequencies $\omega_k^2$) are given by:

$$\omega_1\omega_2\omega_3 = \overset{o}{\omega}_o^3, \qquad \omega_1^2\langle x^2\rangle = \omega_2^2\langle y^2\rangle = \omega_3^2\langle z^2\rangle, \tag{20}$$

where $\overset{o}{\omega}_o$ is given in the paragraph immediately following Eq (4). A convenient parametrization of the oscillator potential frequencies that satisfies the constant volume condition in Eq (20) is [4]:

$$\omega_1 = \omega_o\left[1-\frac{2}{3}\varepsilon\cos\left(\gamma+\frac{2\pi}{3}\right)\right], \quad \omega_2 = \omega_o\left[1-\frac{2}{3}\varepsilon\cos\left(\gamma-\frac{2\pi}{3}\right)\right], \quad \omega_3 = \omega_o\left(1-\frac{2}{3}\varepsilon\cos\gamma\right), \tag{21}$$

$$\omega_o^{-3} = \overset{o}{\omega}_o^{-3}\left[1-\frac{2}{3}\varepsilon\cos\left(\gamma+\frac{2\pi}{3}\right)\right]\cdot\left[1-\frac{2}{3}\varepsilon\cos\left(\gamma-\frac{2\pi}{3}\right)\right]\cdot\left(1-\frac{2}{3}\varepsilon\cos\gamma\right), \tag{22}$$

where $\varepsilon$ is a measure of the overall deformation of the nuclear shape, and $\gamma$ is a measure of the axial asymmetry of the nuclear shape, so that for axial symmetry $\gamma = 0$.

## 2.1 Particle occupation configuration for triaxial rotation

For a triaxial nuclear state, $\omega_1 \neq \omega_2 \neq \omega_3$, and a single-particle state $|\nu\rangle = |N, n_1, n_2, n_3\rangle = |n_1\rangle|n_2\rangle|n_3\rangle$ can be labelled by the well-defined the three decoupled oscillator phonon occupation quantum numbers $n_1, n_2, n_3$ along the three coordinate axes $x, y$, and $z$, and by the total quantum number $N = n_1 + n_2 + n_3$. For an s-d shell nucleus such as $^{20}$Ne, the possible single-particle states are chosen from the list:

$$(000),(100),(010),(001),(200),(020),(002),(110),(101),(011) \tag{23}$$

For a given initial value of the deformation set $(\varepsilon, \gamma)$ in the iteration process, the single-particle energies of the oscillator states in Eq (12) are determined and the lowest-energy states in Eq (23) are populated with four nucleon in each state and subject to the total angular momentum $\hbar I$ to obtain the nuclear state at $(\varepsilon, \gamma)$. A new value of $(\varepsilon, \gamma)$ is determined using Eqs (20) and (21) and this process is repeated until convergence is achieved.

## 2.2 Particle-occupation configuration for axially symmetric rotation

For an axially symmetric nuclear state, $\omega_1 = \omega_2 \neq \omega_3$ and $\gamma = 0$, the z axis is the symmetry axis, and hence $j_3$ is well-defined and the eigenstates can be labelled as follows [2,3,4]: $|N n_3 \Lambda \bar{\Omega}\rangle$ where $\Lambda$ is the $l_3$ quantum number and $\bar{\Omega}$ is $j_3 = l_3 + s_3$ quantum number and the quantum of $s_3$ is $\pm \frac{1}{2}$. Then, $N = n_3 + n_*$ where $n_* \equiv n_1 + n_2$, and $|\Lambda| = n_*, n_* - 2, n_* - 4, \ldots 0$ or $1$. For an s-d shell, we obtain the following axis-symmetric Nilsson single-particle states:

$$[000]_{N=0}, \left[101 \tfrac{1}{2}\right]_{N=1, n_1=1}, \left[101 \tfrac{3}{2}\right]_{N=1, n_2=1}, \left[110 \tfrac{1}{2}\right]_{N=1, n_3=1}, \left[200 \tfrac{1}{2}\right]_{N=2, n_*=2, n_1=2, n_3=0},$$
$$\left[202 \tfrac{5}{2}\right]_{N=2, n_*=2, n_2=2, n_3=0}, \left[220 \tfrac{1}{2}\right]_{N=2, n_*=0, n_3=2}, \left[220 \tfrac{1}{2}\right]_{N=2, n_*=0, n_3=2}, \quad (24)$$
$$\left[202 \tfrac{3}{2}\right]_{N=2, n_*=2, n_1=1, n_3=0}, \left[211 \tfrac{1}{2}\right]_{N=2, n_*=1, n_1=1, n_3=1}, \left[211 \tfrac{3}{2}\right]_{N=2, n_*=1, n_2=1, n_3=1}$$

Fig 1 shows the Nilsson single-particle energies and the crossing of the energy levels versus the deformation parameter $\varepsilon$ for axially-symmetric (i.e., for $\gamma = 0$) s-d shells. The variation of the energies with $\varepsilon$ is similar to that predicted using a numerical solution method in references [1-4].

For a given initial value of $\varepsilon$, the single-particle energies of the spin-orbit oscillator states in Eq (12) are determined and the lowest-energy single-particle states in Eq (24) are populated with four nucleon in each single-particle state to obtain the nuclear state at $\varepsilon$. A new value of $\varepsilon$ is determined using Eqs (20) and (21). This process is repeated iteratively until convergence is achieved.

## 3. Nilsson-CCRM3 predicted rotation, deformation, and yrast energies in $^{20}$Ne

We have used the algebraic Nilsson-CCRM3 model described in Sections 2 and 2.1 to analyze triaxial rotation in $^{20}$Ne nucleus and provide possible explanations for the observed rotational phenomena. Table 1 and Figs 2 to 10 show the results.

The ground state of $^{20}$Ne is axially symmetric and so $\gamma = 0$. We have calculated the minimum ground-state energy $E_o$ and deformation $\varepsilon_o$ to be respectively 51 (in units of $\hbar \overset{o}{\omega}_o$) and 0.26. This $E_o$ and deformation are not accurate because our calculation ignores residual interactions, higher multiple moments, shell energy, and self-consistency (refer to [4,9]). In particular, the ground state quadrupole moment is not accurate as a constraint HF calculation may indicate. Therefore, we have chosen $E_o$ such that the excitation energy $\Delta E_2 \equiv (E_2 - E_o) \times 13.05$ *MeV* at $I = 2$ is as close as possible to the measured excitation energy. This choice fixes $E_o = 50.2$ (in units of $\hbar \overset{o}{\omega}_o$). Using this ground-state energy, we present, in Table 1 and in Figs 2 to 10, the algebraic Nilsson-CCRM3-predicted ground-state rotational-band yrast (i.e. the lowest energy for a given angular momentum $I$) excited-state energy $E_I$, the excitation energy $\Delta E_I \equiv (E_I - E_o) \times 13.05$ *MeV*, and the rotation types and nature for $I = 2, 4, 6,$ and $8$.

Figs 2 to 9 show the variation, with iteration number, in the algebraic Nilsson-CCRM3-predicted yrast-state energy and deformation parameters $\varepsilon$ and $\gamma$. In Figs 2 to 9, self-consistency in the model variables is achieved only when the difference between input and output parameter (i.e., $\varepsilon$ and $\gamma$) values is acceptably small, i.e., the iteration on the deformation parameters and energy yields nearly a converged value, and the energy value is the lowest.

For $I = 2$ and $6$, Figs 2 and 6 show that the yrast excited-state energies $E_2$ and $E_6$ vary steadily with the iteration number, eventually attaining converged values.

However, for $I = 4$ and $8$, Figs 4 and 8 show that the yrast excited-state energies $E_4$ and $E_8$ vary periodically or cyclically with the iteration number (i.e., $\varepsilon$ and $\gamma$ values). For $I = 4$, $E_4$ in Fig 4 exhibits persistent oscillations between $E_4 = 51$ and $52.5$ (in units of $\hbar \overset{o}{\omega}_o$). Fig 4 shows that the deformation parameter set ($\varepsilon$ and $\gamma$) exhibits persistent oscillations between (0.1472,0.4802) and (0.1755,0.6753). These oscillations are generated by the self-consistency conditions and alternate crossings of the same two single-particle levels in certain ranges of $\varepsilon$ and $\gamma$, thereby coupling the energy states at 51 and 52.5. We have decoupled the lower and upper energy states by using only the lower-energy-state deformation parameter set (0.1755,0.4802) in the iteration steps starting at the iteration step number 115 in Figs 4 and 5. This decoupling reduces the lower-state energy from 51 to 50.

For $I=8$, $E_8$ and the deformation parameters in Figs 8 and 9 undergo cyclic variations with the iteration number. These cyclic variations eventually decay to steady variation (because of the crossing of different single-particle energy levels), and converge to lower energy values, as observed in Figs 8 and 9. The reduction in energy at $I=8$ in $^{20}$Ne seems to be somewhat similar to that predicted in [refer to footnote 1], which is caused by a feedback between the angular momentum and the microscopic MSCRM3 (described in Section 1) angular velocity resulting in the quenching of planar rotation to uniaxial rotation.

Fig 10 shows measured yrast-state excitation energy versus $I$, and also the excitation energy $\Delta E_I \equiv (E_I - E_o) \times 13.05\, MeV$ predicted by the algebraic Nilsson-CCRM3 model in this article and that predicted in articles [4,9]. Each of the three plots in Fig 10 shows a reduction in the excitation energy at $I=4$ and 8. Table 1 and Fig 10 show that the excitation energy predicted in this article is in a much better agreement with the measured excitation energy than that predicted in [4,9].

As may be deduced from Figs 8 and 9, oscillatory or cyclic variations in the ground-state band members $I=4$ and $I=8$ may eventually cause in the observed reduction of the steady values of energies of $I=4$ and $I=8$ states relative to those of $I=2$ and $I=6$, as suggested in [2, page 96].

Table 1: Measured and Nilsson-CCRM3 predicted excited-state and excitation energies,
nuclear shape and rotation types

| angular momentum $I$ | excited-state energy $E_I$ in units of $\hbar \overset{o}{\omega}_o$ | excitation energy $\Delta E_I \equiv (E_I - E_o)$ $\times 13.05\ MeV$ | deformation $\varepsilon$ | axial asymmetry parameter $\gamma$ rad | measured $\Delta E_I$ $MeV$ | comments |
|---|---|---|---|---|---|---|
| 2 | 50.2931 $E_2$ in Figs 2 and 3 does not exhibit oscillations | 1.215 | 0.5058 | 0.1045 | 1.6335 | Triaxial prolate shape close to axial symmetry, and nearly planar rotation in x-y plane similar to that in footnote 1 |
| 4 | 50.5007 $E_4$ in Fig 4 and 5 exhibits oscillations between 51 and 52.5 values with iteration # | 3.9241 | 0.543 | 0.167 | 4.247 | Highly prolate and nearly axially symmetric shape about z axis, and triaxial rotation mostly about x axis along which quad moment is smallest, whereas footnote 1 predicts prolate shape with uniaxial rotation about x axis |
| 6 | 50.8545 $E_6$ value in Figs 6 and 7 is very close to the other predicted $E_6$ values but it is the most self-consistent | 8.5412 | 0.7595 | 0.732 | 8.775 | Triaxial highly prolate shape, and triaxial rotation mostly about x axis, whereas footnote 1 predicts uniaxial rotation about the x axis |
| 8 | 51.0884 $E_8$ in Figs 8 and 9 exhibits cyclic variations with iteration # | 11.5936 | 0.6734 | 1.0472 (60º) | 11.948 | almost oblate axially symmetric about x axis and triaxial rotation with highest rotation about x axis |

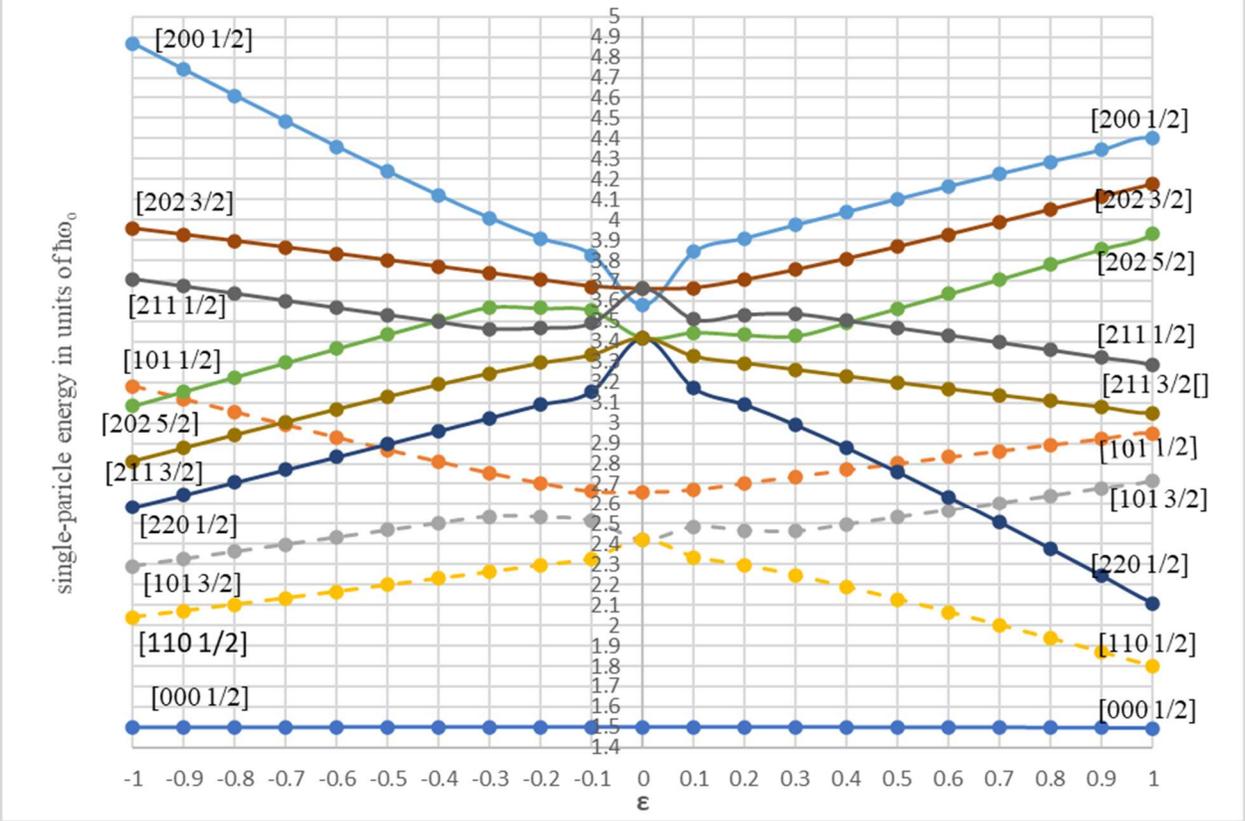

Fig. 1: Single-particle Nilsson energy $e_{\upsilon k}$ in units of $\hbar\omega_o$ versus axially-symmetric oscillator deformation parameter $\varepsilon$

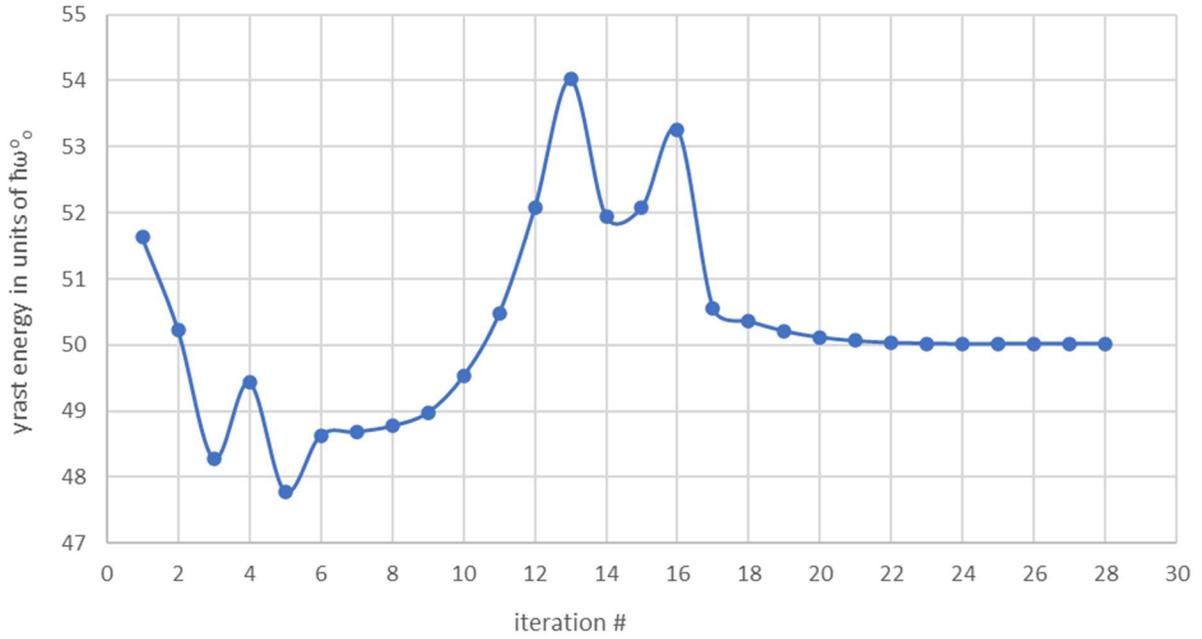

Fig 2: CCRM3-predicted cranked yrast-state energy at $I=2$ in units of $\hbar\omega^o_o$ versus iteration # for $^{20}$Ne nucleus

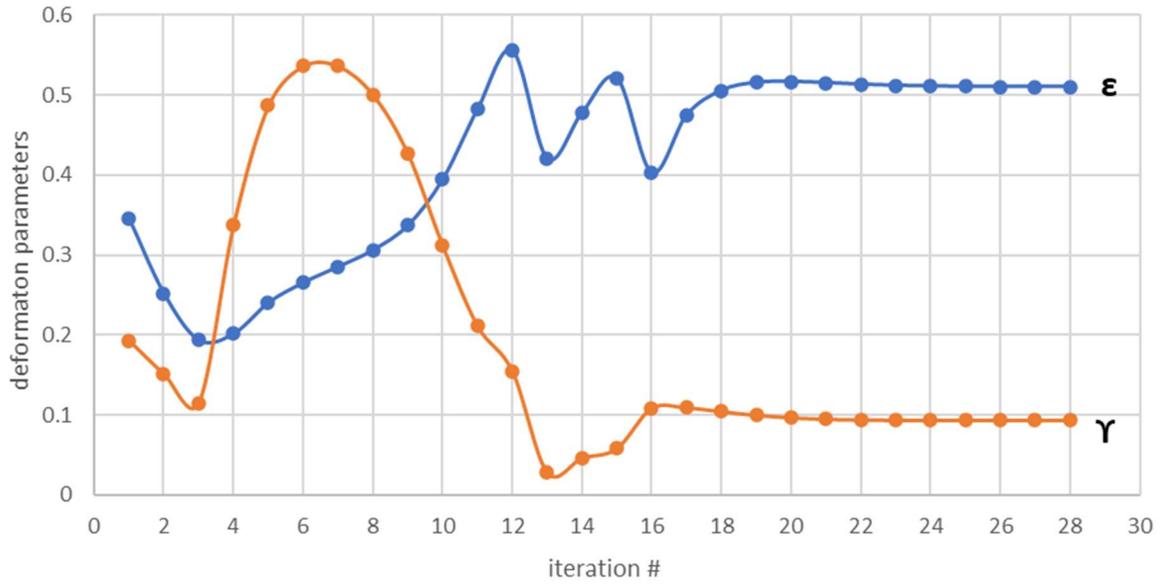

Fig 3: CCRM3-predicted values of deformation parameters $\varepsilon$ and $\Upsilon$ at $I=2$ versus iteration # for $^{20}$Ne nucleus

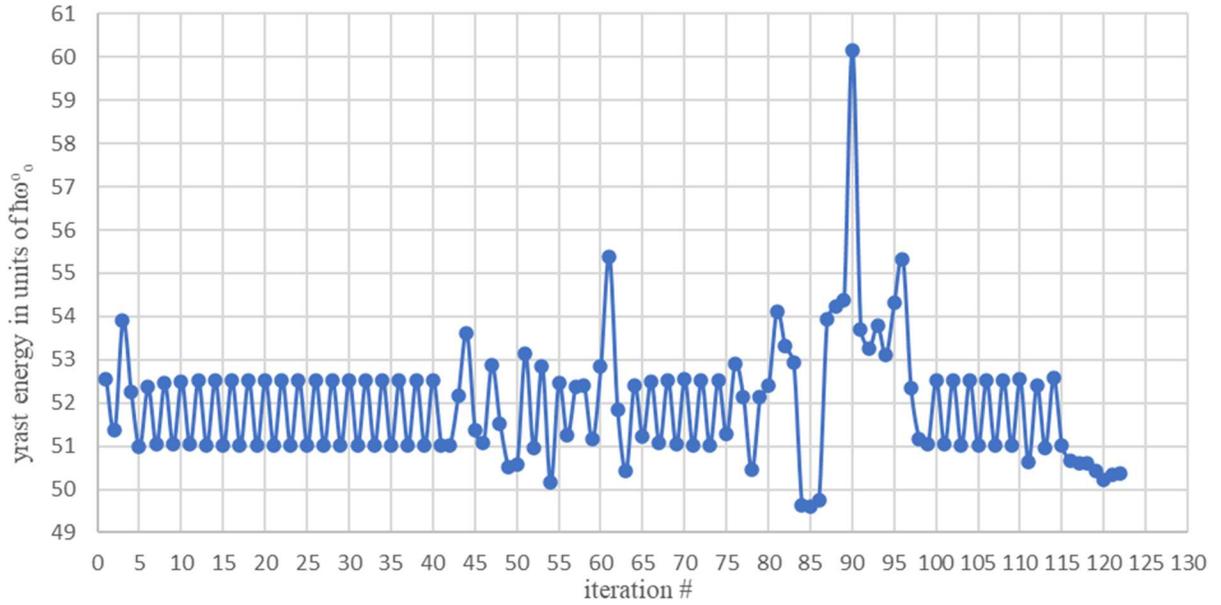

Fig 4: CCRM3-predicted cranked yrast-sate energy $E_4$ in units of $\hbar\omega^o_o$ at $I=4$ versus iteration number for $^{20}$Ne nucleus

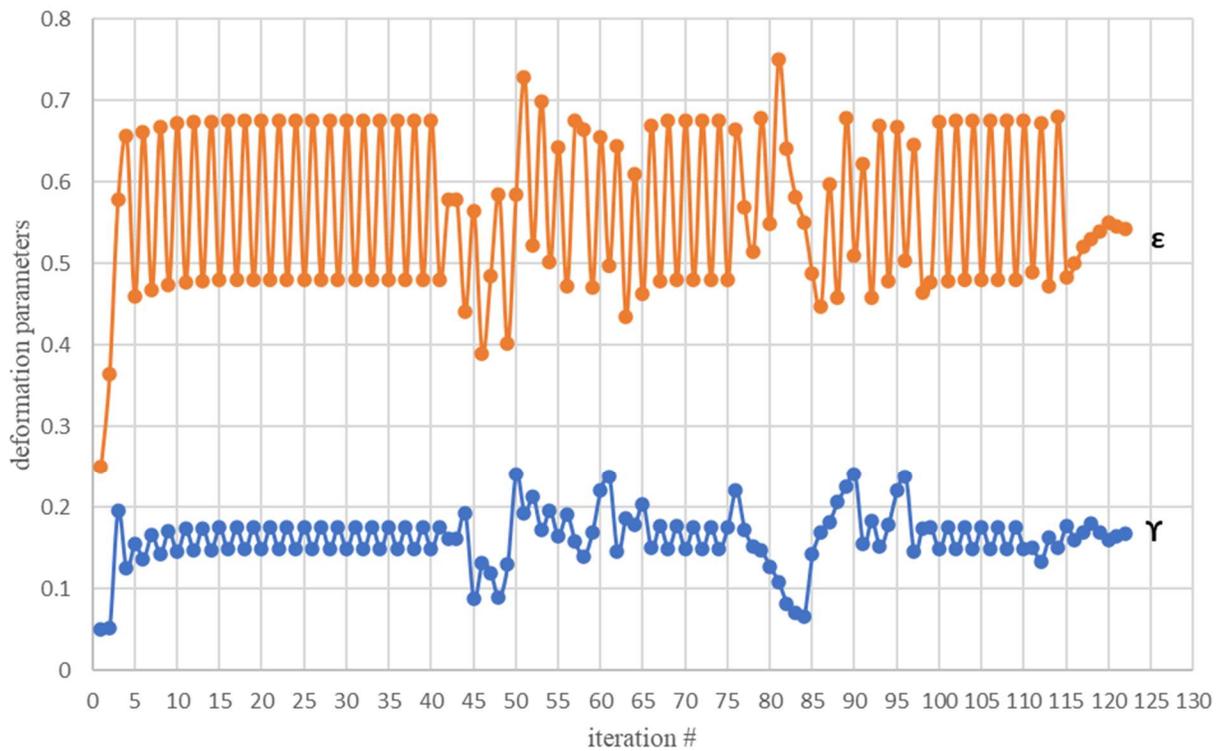

Fig 5: CCRM3-pedicted values of deformation parameters $\varepsilon$ and $\Upsilon$ at $I=4$ versus iteration # for $^{20}$Ne nucleus

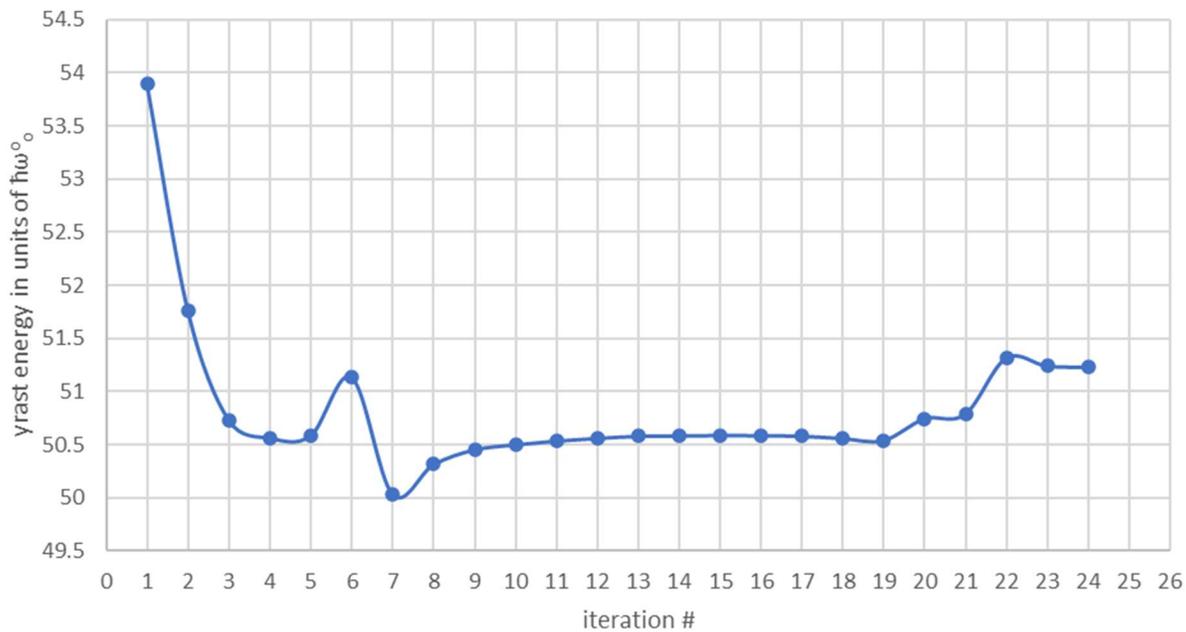

Fig 6: CCRM3-predicted yrast-state energy in units of $\hbar\omega°_o$ at $I=6$ versus iteration # for yrast-state $I=6$ for $^{20}$Ne nucleus

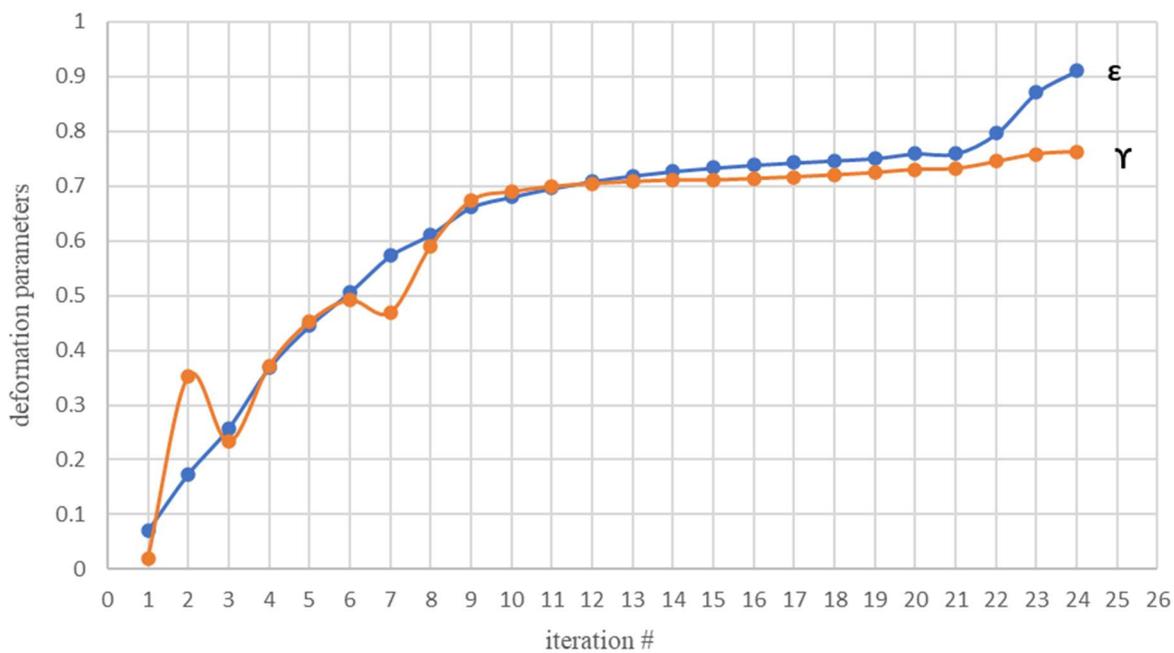

Fig 7: CCRM3-predicted deformation parameters $\varepsilon$ and $\Upsilon$ versus iteration # for yrast state at $I=6$ for $^{20}$Ne nucleus

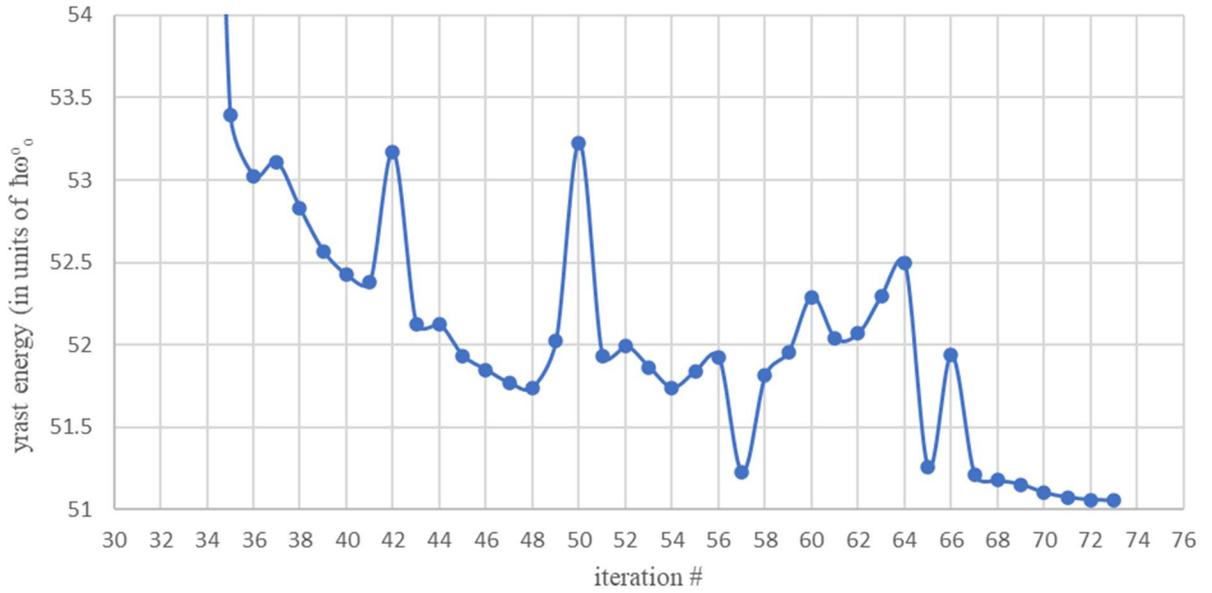

Fig 8: CCRM3-predicted cranked yrast energy $E_\delta$ in units of $\hbar\omega°_o$ at $I=8$ versus iteration # for $^{20}$Ne nucleus

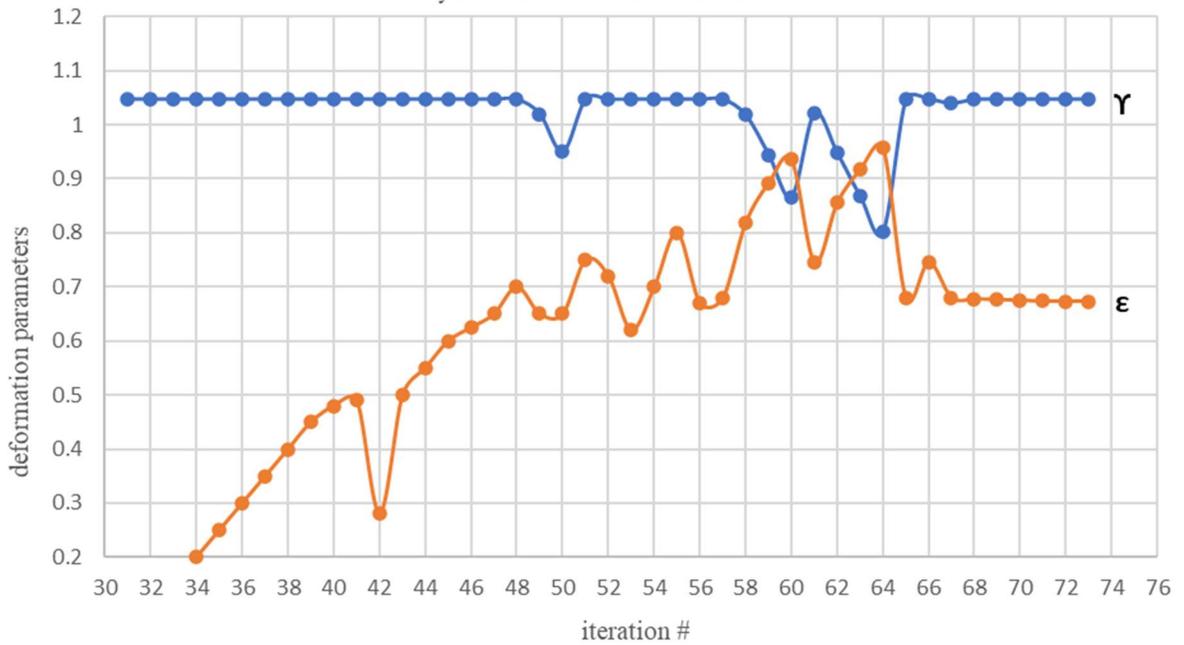

Fig 9: CCRM3-predicted values of deformation parameters ε and ϒ versus iteratrion # for yrast state at $I=8$ for $^{20}$Ne nucleus

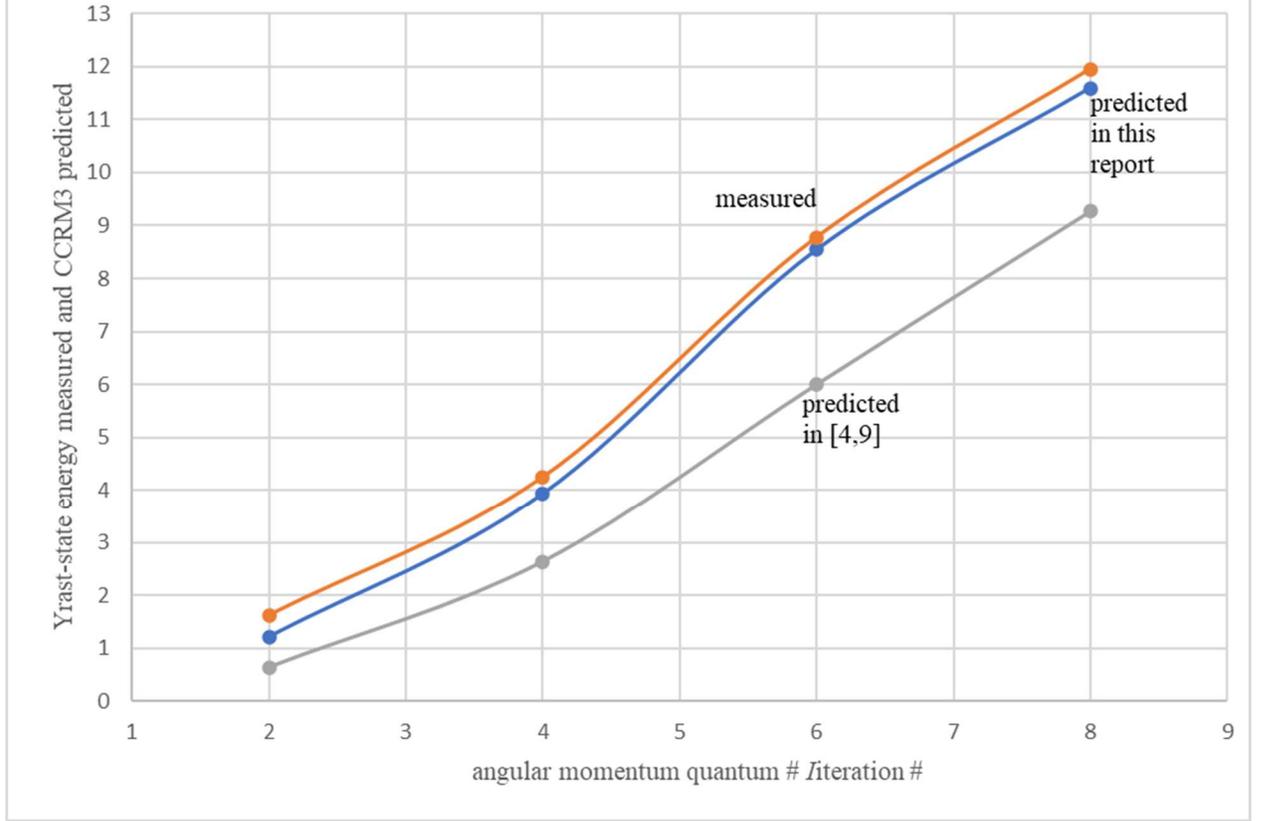

Fig 10: Measured and CCRM3-predicted yrast-state energy versus *I* in this report and in references [4,9]

## 4. Summary, conclusions, and future plans

Most analyses of nuclear uniaxial and triaxial collective rotations use the conventional cranking model with a constant angular velocity (CCRM3) and with the Nilsson deformed harmonic oscillator mean-filed potential energy, which includes the nuclear spin-orbit interaction to make it more representative of the situation in real nuclei. In [4,9], the Nilsson-CCRM3 Schrodinger equation was solved self-consistently using a numerical method [4,9]. In this article, we use an algebraic method (used previously in [ refer to footnote 1] for the pure oscillator potential) to solve iteratively the self-consistent Nilsson-CCRM3 Schrodinger equation. In the application of the algebraic Nisson-CCRM3 to $^{20}$Ne, we have not used, as in all previous analyses, pairing correlations because they are not significant in the light nucleus $^{20}$Ne.

The application of the Nilsson-CCRM3 algebraic model to the $^{20}$Ne predicts ground-state rotational-band yrast excitation energies at angular momenta *I* =2,4,6, and 8 that are in a much better agreement with the measured energies than those predicted in [4,9] using a numerical solution method.

The algebraic Nilsson-CCRM3 model predicts oscillations, with the iteration steps, in the excited-state energy at *I* =4, which are generated by the crossings of the single-particle energy levels, such as those seen in Fig 1 and by the self-consistency conditions in Eqs (20) and (21). These oscillations indicate that the energy consists of two coupled states, a state at lower energy and another at higher energy. By using, in the iteration, only the values of the deformation parameters related to the lower energy component, we isolate the yrast state and obtain its energy at slightly lower energy than the lower-energy component in the coupled system. For *I* =8, the energy $E_8$ and the deformation parameters exhibit cyclic variations with the iteration number. These cyclic variations eventually decay to a steady variation (because of different single-particle energy level crossings), and converge to a lower energy. These results seem to support the Bohr-Mottelson's suggestion [2, page 96] that the oscillations result in the observed reduction in the *I* =4 and 8 energies.

It is noted that the reduction in the excitation at $I=8$ in $^{20}$Ne discussed above appears to be somewhat similar to the reduction in energy at $I=8$ in $^{20}$Ne predicted in [refer to footnote 1] using MSCRM3, where the reduction in energy is caused by a feedback between the angular momentum and the microscopic MSCRM3 (described in Section 1) angular velocity resulting in the quenching or transition from planar rotation to uniaxial rotation.

We plan to use the algebraic Nilsson-CCRM3 model where the angular velocity is replaced by the microscopic angular velocity in MSCRM3, described above and in Section 1 and developed in [refer to footnote 1], in order to obtain a better understanding of the variation of rotational-band yrast excitation energy with angular momentum $I$.